\documentclass[pra,aps,twocolumn,showpacs,floatfix]{revtex4}
\usepackage{amsfonts}
\usepackage{amssymb}
\usepackage{hyperref}
\usepackage{graphicx}
\usepackage{dcolumn}
\usepackage{bm,amsmath,verbatim}
\usepackage{mathrsfs}
\usepackage{color}

\def\be{\begin{equation}}
\def\ee{\end{equation}}
\def\bea{\begin{eqnarray}}
\def\eea{\end{eqnarray}}
\def\bse{\begin{subequations}}
\def\ese{\end{subequations}}

\def\be{\begin{eqnarray}}
\def\ee{\end{eqnarray}}

\begin{document}

\title{Type-II Weyl Points in Three-Dimensional Cold Atom Optical Lattices}
\author{Yong Xu}
\email{yongxuph@umich.edu}
\author{L.-M. Duan}
\email{lmduan@umich.edu}
\affiliation{Department of Physics, University of Michigan, Ann Arbor, Michigan 48109, USA}

\begin{abstract}
Topological Lifshitz phase transition characterizes an abrupt change of the topology
of the Fermi surface through a continuous deformation of parameters. Recently,
Lifshitz transition has been predicted to separate two types of Weyl points:
type-I and type-II (or called structured Weyl points), which has attracted considerable attention in various fields. Although
recent experimental investigation has seen a rapid progress on type-II Weyl points,
it still remains a significant challenge to observe their characteristic Lifshitz transition.
Here, we propose a scheme to realize both type-I and type-II Weyl points in three-dimensional
ultracold atomic gases by introducing an experimentally feasible configuration based on
current spin-orbit coupling technology. In the resultant Hamiltonian, we find
three degenerate points: two Weyl points carrying a Chern number $-1$ and a four-fold degenerate point
carrying a Chern number $2$. Remarkably, by continuous tuning of
a convenient experimental knob, all these degenerate points can transition from
type-I to type-II, thereby providing an ideal platform to study different types of Weyl
points and directly probe their Lifshitz phase transition.
\end{abstract}
\maketitle

\section{Introduction}

Distinct from conventional phase transitions driven by spontaneous symmetry breaking,
Lifshitz phase transition is driven by an abrupt change of the topology
of the Fermi surface~\cite{Lifshitz}. Recently, it has been predicted in three-dimensional (3D)
condensed matter materials that type-I Weyl fermions (i.e., Weyl points)~\cite{volovik,Murakami2007NJP,Wan2011prb,Vafek2014}
can transition to type-II~\cite{Soluyanov2015Nature} (or called structured Weyl points~\cite{Yong2015PRL}) through Lifshitz transition~\cite{Soluyanov2015Nature,Yong2015PRL}.
Although type-I Weyl fermions~\cite{Weyl} were initially predicted in particle physics,
type-II Weyl fermions may not be allowed to exist there due to Lorentz symmetry,
which is absent in condensed matter materials.
Such Lifshitz transition is perceived as the change of a type-I Weyl point's single point Fermi surface
to a type-II's open Fermi surface consisting of particle and hole pockets along
with a touching point~\cite{Yong2015PRL,Soluyanov2015Nature}.
While remarkable experimental and theoretical progress has been reported recently on
type-II Weyl semimetals and their properties~\cite{Yong2015PRL,Soluyanov2015Nature,Bergholtz2015PRL,
Volovik2016,Yang2016,Udagawa2016,Tiwari2016,Beenakker2016,
Vernevig2016M, Chang2016NC, Koepernik2016PRB,
Liang2016, Kaminskyi2016, Xu2016, Deng2016, Chen2016, Zhou2016,
Shi2016,Bruno2016}, it remains a significant
challenge in solid-state materials to observe the characteristic Lifshitz
phase transition between type-I and type-II Weyl points.

Ultracold atomic gases provide an ideal platform to observe the Lifshitz
transition because of their high controllability. And recent experiments on
one-dimensional and two-dimensional (2D) spin-orbit
coupling~\cite{Lin2011Nature, Jing2012PRL, Zwierlen2012PRL, PanJian2012PRL,
Qu2013PRA, Jing2015Nature, Jing2015,Shuai2015,Lev2016,Ketterle2016} in ultracold atomic gases
have further paved the way for discovering novel
topological quantum states \cite{Bloch2012,SpielmanReview,ZollerReview}. Although type-I
and type-II Weyl points were proposed in quasiparticle spectra of spin-orbit
coupled Fermi superfluids~\cite{Gong2011prl,
Sumanta2013PRA,Yong2014PRL,Dong2014arXiv,Yong2015PRL,Xu2015Review}, realization of such
superfluids is still a big challenge with current experimental technology~%
\cite{Jing2012PRL,Zwierlen2012PRL,Jing2015Nature,Jing2015}. A more feasible
experimental scheme is to realize Weyl points in the single-particle spectra
without the need of low-temperature Fermi superfluids. While some proposals have
been made concerning type-I Weyl points \cite{Anderson2012PRL,
Jiang2012PRA,Ganeshan2015PRB,Tena2015RPL,Law2015}, the scheme to realize
both type-I and type-II Weyl points and their Lifshitz transition in
the single-particle spectra of cold atoms is still lacking and highly desired.

In this paper, we propose a novel 3D model to realize both type-I and type-II Weyl
points in the single-particle spectra of cold atoms in 2D optical lattices.
In this model, we find two Weyl points carrying a Chern number -1 and a four-fold
degenerate point carrying a Chern number 2,
protected by 2D pseudo-time-reversal (2D-PTRS), 2D inversion (2D-IS), and combined rotational
symmetries. By continuous tuning of a Zeeman field, all these degenerate points
can experience the Lifshitz transition from type-I to type-II.
Furthermore, we find a laser-atom coupling configuration to implement the model
based on the current experimental technology that realizes the required spin-orbit
coupling. Due to the controllability of the Zeeman field through continuous
tuning of a convenient experimental knob, this system offers a unique
opportunity to observe the characteristic topological Lifshitz quantum
phase transition between type-I and type-II Weyl points.

\section{Model Hamiltonian}

We start by briefly reviewing the concept of type-I and type-II Weyl points
that are described by an effective Hamiltonian $H_{W}=v_{0}k_{z}+\sum_{\mu
=x,y,z}v_{\mu }k_{\mu }\sigma _{\mu }$ \cite{Yong2015PRL,Soluyanov2015Nature}%
, where $k_{\mu }$ ($\sigma _{\mu }$) denote momenta (Pauli matrices) and $%
v_{0},v_{\mu }$ are real parameters. Its energy spectrum is given by $E_{\pm
}(\mathbf{k})=v_{0}k_{z}\pm \sqrt{\sum_{\mu =x,y,z}v_{\mu }^{2}k_{\mu }^{2}}$
with $\pm $ labeling particle and hole bands. When $|v_{0}|<|v_{z}|$, the
energy of the particle (hole) band is positive (negative), except at the
touching point where the energy vanishes. This touching point is dubbed
type-I Weyl point. When $|v_{0}|>|v_{z}|$, in certain regions, the energy
goes negative for the particle band and positive for the hole band, leading
to an open Fermi surface besides a touching point at $E_{\pm }(\mathbf{k}=0)=0$%
. This structure is dubbed type-II Weyl point.

To realize both type-I and type-II Weyl points with cold atoms, we consider the
following Hamiltonian that describes atoms in 2D optical lattices
\begin{equation}
H^{\prime }=\frac{\mathbf{p}^{2}}{2m}+\sum_{\nu =x,y}V_{\nu }\sin
^{2}(k_{L\nu }r_{\nu })+h_{z}\sigma _{z}+V_{SO},
\label{HC}
\end{equation}%
where $\mathbf{p}=-i\hbar \nabla $ is the momentum operator, $m$ is the mass
of atoms, $V_{\nu }$ ($\nu =x,y$) denote the strength of optical
lattices with the period being $a_{\nu }=\pi /k_{L\nu }$ along the $\nu $
direction, $h_{z}$ is the Zeeman field, $\sigma _{\nu }$ are Pauli matrices
for spins, and $V_{SO}$ is a laser-induced spin-orbit coupling term taking
the form
\begin{equation}
V_{SO}=\Omega _{SO}(M_{x}+iM_{y})e^{ik_{Lz}r_{z}}|\uparrow \rangle
\left\langle \downarrow \right\vert +\text{H.c}.
\end{equation}%
with $M_{x}=\sin (k_{Lx}r_{x})\cos (k_{Ly}r_{y})$, $M_{y}=\sin
(k_{Ly}r_{y})\cos (k_{Lx}r_{x})$, $\Omega _{SO}$ proportional to the
laser strength, and $k_{L\nu }$ with $\nu=x,y,z$ determined by lasers'
wave vector along the $\nu$ direction. Later, we will describe the laser configuration that
directly realizes the Hamiltonian (1), in particular the spin-orbit coupling
term $V_{SO}$. Employing a unitary transformation with
$U=e^{-ik_{Lz}r_{z}/2}|\uparrow\rangle\langle\uparrow|+e^{ik_{Lz}r_{z}/2}|\downarrow\rangle\langle\downarrow|$
yields $H=UH^\prime U^{-1}$ with
\begin{equation}
H=\frac{\hbar ^{2}k_{z}^{2}}{2m}+\tilde{h}_{z}\sigma _{z}+H_{2D},
\end{equation}%
where $k_{z}=p_{z}/\hbar $, $\tilde{h}_{z}=\hbar ^{2}k_{L{z}}k_{z}/(2m)+h_{z}
$, and the 2D\ Hamiltonian $H_{2D}$ in the $(x,y)$ plane is expressed as
\begin{eqnarray}
H_{2D} &=&\sum_{\nu =x,y}\left[ \frac{p_{\nu }^{2}}{2m}+V_{\nu }\sin
^{2}(k_{L\nu }r_{\nu })\right]   \notag \\
&+&\left[ \Omega _{SO}(M_{x}+iM_{y})|\uparrow \rangle \left\langle
\downarrow \right\vert +\text{H.c.}\right] .
\end{eqnarray}

\begin{figure}[t]
\includegraphics[width=3.4in]{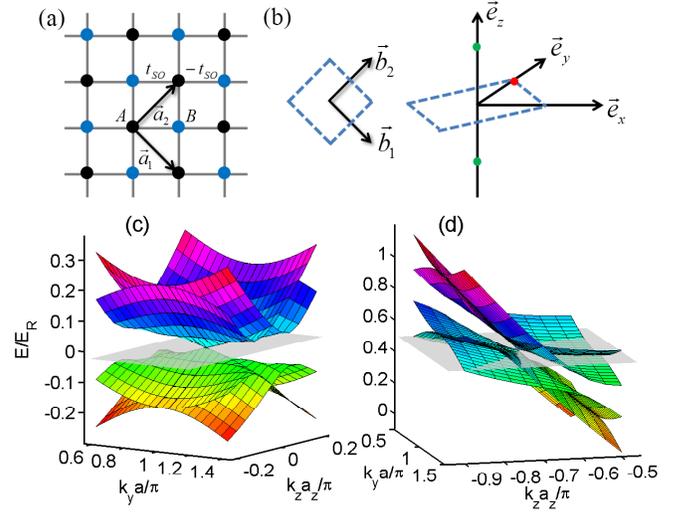}
\caption{(Color online) (a) Lattice structure in the $(x,y)$ plane, where
each unit cell is made up of A and B sites and $\mathbf{a}_1=a\mathbf{e}_x-a%
\mathbf{e}_y$ with $a=a_x=a_y$ and $\mathbf{a}_2=a\mathbf{e}_x+a\mathbf{e}_y$
are unit vectors. (b) The first brillouin zone denoted by the dashed box. $%
\mathbf{b}_1=\protect\pi(\mathbf{e}_x-\mathbf{e}_y)/a$ and $\mathbf{b}_2=%
\protect\pi(\mathbf{e}_x+\mathbf{e}_y)/a$ are reciprocal unit vectors. Green
[${\bf k}^{W \pm}=(k_x^{W \pm}a,k_y^{W \pm}a,k_z^{W \pm}a_z)=(0,0,2m\pi(\pm 4\bar{t}-h_z)/(\hbar^2 k_{Lz}^2)$]
and red solid [${\bf k}^{W0}=(k_x^{W0}a,k_y^{W0}a,k_z^{W0}a_z)=(0,\pi,-2m\pi h_z/(\hbar^2 k_{Lz}^2))$]
circles denote the position where Weyl points exist in the momentum space.
Here, $a_z=\protect\pi/k_{Lz}$. (c)(d) Spectra
near a type-I and type-II four-fold degenerate point with respect to $(k_ya,k_za_z)$ for $%
k_xa=0$ (see the spectra of the continuous model in Appendix A
). In (d), the gray plane (i.e., $E=E_W$ with $E_W$ being the energy at the
degenerate point) intersects both the particle and hole bands in addition to
the degenerate point, implying it is type-II. Here,
$E_R=\hbar^2k_R^2/2m$ with $k_R$ being the
absolute value of lasers' wave vector. }
\label{Fig1}
\end{figure}

To see how the Weyl points emerge in this model, we discretize $H_{2D}$ and
study its physics in the tight-binding model (see Appendix A
for details of discretization). The tight-binding form of $H$ can be written
as [let us first neglect $\hbar ^{2}k_{z}^{2}/(2m)$ term and focus on type-I
Weyl points]
\begin{eqnarray}
H_{TB}&&=\sum_{k_{z}}\sum_{{\bf x}}[ \tilde{h}_{z}\hat{c}_{k_{z},{\bf x}}^{\dagger }\sigma_z\hat{c}_{k_{z},{\bf x}}
+\sum_{\nu=x,y} (-t_{\nu}\hat{c}_{k_{z},{\bf x}}^{\dagger }\hat{c}_{k_{z},{\bf x}+{\bf g}_\nu}
\nonumber \\
&&+(-1)^{j_x+j_y}t_{SO\nu}\hat{c}_{k_{z},{\bf x}}^{\dagger }\sigma_\nu\hat{c}_{k_{z},{\bf x}+{\bf g}_\nu}+\text{H.c.}) ],
\end{eqnarray}
where ${\bf g}_\nu=a_\nu{\bf e}_\nu$; $\hat{c}^\dagger_{k_{z},{\bf x}}=(
\begin{array}{cc} \hat{c}^\dagger_{k_{z},{\bf x},\uparrow} & \hat{c}^\dagger_{k_{z},{\bf x},\downarrow}
\end{array}
)$ with $\hat{c}_{k_{z},{\bf x},\sigma }^{\dagger }$ ($\hat{c}_{k_{z},{\bf x},\sigma }$) being the
creating (annihilating) operator and ${\bf x}=j_{x}a_x{\bf e}_x+j_{y}a_y{\bf e}_y$;
$t_\nu$ and $t_{SO\nu}$ denote the tunneling and spin-orbit coupling
strength along the $\nu$ direction.

Different from the well-known 2D Chern insulator~\cite{Qi2008,Huges2016},
this Hamiltonian involves the position dependent spin-orbit coupling, and we thus
need to choose a unit cell consisting of two sites: A and B [as shown in
Fig.~\ref{Fig1}(a)] and the Hamiltonian in the momentum space in
the new basis $\Psi (\mathbf{k})^{T}$ with $\Psi (\mathbf{k})=(%
\begin{array}{cccc}
e^{ik_{x}a_{x}}\hat{A}_{\mathbf{k}\uparrow } & e^{ik_{x}a_{x}}\hat{A}_{%
\mathbf{k}\downarrow } & \hat{B}_{\mathbf{k}\uparrow } & \hat{B}_{\mathbf{k}%
\downarrow }%
\end{array}%
)$
reads
\begin{equation}
H_{TB}(\mathbf{k})=\tilde{h}_{z}\sigma _{z}-h_{t}\tau _{x}+\tau
_{y}(-d_{x}\sigma _{x}+d_{y}\sigma _{y}),  \label{HTBk}
\end{equation}%
where $h_{t}=2\sum_{\nu =x,y}t_{\nu }\cos (k_{\nu }a_{\nu })$, $%
d_{x}=2t_{SOx}\sin (k_{x}a_{x})$ and $d_{y}=-2t_{SOy}\sin (k_{y}a_{y})$; $%
\tau $ are Pauli matrices acting on A, B sublattices. In the $(k_{x},k_{y})$
plane, $H_{TB}(\mathbf{k}_{\perp },k_{z})$ ($\mathbf{k}_{\perp }=k_{x}\mathbf{e}%
_{x}+k_{y}\mathbf{e}_{y}$) respects 2D-IS: $\tau _{x}H_{TB}(\mathbf{k})\tau _{x}=H_{TB}(-%
\mathbf{k}_{\perp },k_{z})$ and when $\tilde{h}_{z}=0$, 2D-PTRS: $%
\mathcal{T}H_{TB}(\mathbf{k})\mathcal{T}^{-1}=H_{TB}(-\mathbf{k}_{\perp },k_{z})$ with
$\mathcal{T}=i\tau _{x}\sigma _{y}\mathcal{K}$ and $\mathcal{K}$ being the
complex conjugate operator. These two symmetries guarantee that the spectrum
in this specific plane ($\tilde{h}_{z}=0$) is at least doubly degenerate,
implying that the touching point, if exists, is four-fold degenerate. We note
that in the continuous model 2D-IS corresponds to $\mathcal{I}_{C}H%
\mathcal{I}_{C}^{-1}=H(-r_x+\pi /k_{L{x}},-r_y,r_z)=H$ with the
inversion center located at $(\pi /(2k_{L{x}}),0)$ and 2D-PTRS corresponds to $\mathcal{%
T}_{C}H_{2D}\mathcal{T}_{C}^{-1}=H_{2D}$ with $\mathcal{T}_{C}\equiv i%
\mathcal{I}_C\mathcal{P}\sigma _{y}\mathcal{K}$ and $\mathcal{P}H_{2D}%
\mathcal{P}^{-1}=H_{2D}(-r_x,-r_y)$.

Specifically, the eigenvalues of $H_{TB}(\mathbf{k})$ read $E_{\mathbf{k}%
}=\pm \sqrt{d_{\perp }^{2}+(h_{t}\pm \tilde{h}_{z})^{2}}$ with $d_{\perp
}^{2}=d_{x}^{2}+d_{y}^{2}$, which supports the above symmetry analysis
that the energy band at each $\mathbf{k}$ is doubly degenerate without $%
\tilde{h}_{z}$. Clearly, when $d_{x}=d_{y}=0$ and $h_{t}\pm \tilde{h}_{z}=0$%
, there emerge degenerate points. This requires $(k_{x}a_{x},k_{y}a_{y})=(0,\pi )$
or $(0,0)$. In the former case, a single degenerate point appears at $%
k_{z}^{W0}a_{z}=-2m\pi h_{z}/(\hbar ^{2}k_{L{z}}^{2})$ if $t_{x}=t_{y}$
(thus $h_{t}=0$ at this point) as a result of a combined rotational symmetry
(i.e., $U_{4}HU_{4}^{-1}=H$ where $U_{4}=S_{4}C_{4}$ with $S_{4}=e^{i\pi
\sigma _{z}/4}$ and $C_{4}$ being the four-fold rotational operator along $z$
when $V_{x}=V_{y}$ and $k_{L{x}}=k_{L{y}}$) readily achievable in
experiments; in this plane, both 2D-PTRS and 2D-IS are preserved and
this point's degeneracy is therefore four-fold (also seen from the eigenvalues).
The dispersion is linear along all three momenta directions as visually
shown in Fig.~\ref{Fig1}(c). Compared with a Dirac point in a
Hamiltonian with both TRS and IS in 3D~\cite{Huges2016}, in our case,
two similar symmetries (2D-PTRS and 2D-IS) are both respected only in the plane $\tilde{h}_{z}=0$. In
fact, the four-fold degenerate point can be viewed as consisting of two Weyl points with the same Chern
number. To demonstrate this, let us write down the effective Hamiltonian near the point,
which, after a unitary transformation, reads
\begin{equation}
H(\mathbf{q})\sim (v_{z}q_{z}\sigma _{z}+v_{x}q_{x}\sigma
_{x}+v_{y}q_{y}\sigma _{y})\tau _{0},
\end{equation}%
where $v_{z}=\hbar ^{2}k_{L{z}}/2m$, $v_{x}=2t_{SOx}a_{x}$, and $%
v_{y}=-2t_{SOy}a_{y}$; $\tau _{0}$ is a $2\times 2$ identity matrix; the
momenta $\mathbf{q}$ are measured with respect to the degenerate point. It is
clear that each spinor corresponds to a Weyl point with the same chirality~\cite{Yang2015Nano}.

To characterize the topological charge of a degenerate point, we define the first
Chern number
\begin{equation}
C=\frac{1}{2\pi }\sum_{n=1,2}\oint_{\mathcal{S}}{\bm\Omega }_{n}(\bm k)\cdot
d{\bm S},  \label{ChernEq}
\end{equation}%
where the surface $\mathcal{S}$ encloses a considered degenerate point, and ${\bm\Omega }%
_{n}(\bm k)=i\langle \nabla _{\bm k}u_{n}(\bm k)|\times |\nabla _{\bm %
k}u_{n}(\bm k)\rangle $ is the Berry curvature~\cite{XiaoRMP} for the $n$-th
band with $|u_{n}(\bm k)\rangle $ being its wave function. For our
parameters, a direct calculation yields $C=2$ for the above discussed four-fold degenerate point.

In the latter case where $(k_{x}^{W\pm}a_{x},k_{y}^{W\pm}a_{y})=(0,0)$,
two degenerate points occur at $k_{z}^{W\pm
}a_{z}=2m\pi (\pm 4\bar{t}-h_{z})/(\hbar ^{2}k_{L{z}}^{2})$ with $\bar{t}%
=(t_{x}+t_{y})/2$. Their degeneracy is double instead of four-fold (owing to the breaking of
2D-PTRS), and their dispersion is also linear in all three momentum directions. Our
calculation demonstrates $C=-1$ for either of them.
Due to the existence of the energy $\hbar ^{2}{%
k_{z}}^{2}/(2m)$, the energy at these two Weyl points is different except when $%
h_{z}=0$. Moreover, while we consider single-particle physics, a crude estimate using
mean-field analysis suggests that weak repulsive short-range interactions may
shift the locations of Weyl and four-fold degenerate points along $z$ and cause
the transition between the type-I and type-II but not destroy them (see Appendix B).

\begin{figure}[t]
\includegraphics[width=3.4in]{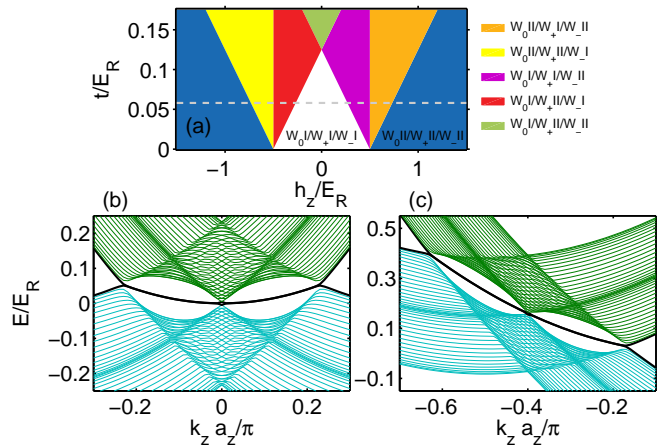}
\caption{(Color online) (a) Phase diagram with respect to $h_z$ and $t$ ($t=t_x=t_y$),
in which $W_\beta {\text{I}}$ ($W_\beta \text{II}$) with $\beta=0,+,-$ represent
the phase with a type-I (type-II) Weyl or four-fold degenerate point located at ${\bf k}^{W\beta}$.
The dashed line shows the case for $t=0.058E_R$.
(a)(b) Spectra as a function of $k_za_z$ for $k_y=0$ under
open boundary condition along the $x$ direction. In (a), $h_z=0$ and the
Weyl points are all type-I; in (b), $h_z=0.4E_R$ and a Weyl point on the
left side is type-II and all others are type-I. The black lines denote
surface states.}
\label{Fig2}
\end{figure}

To illustrate that these type-I Weyl and four-fold degenerate points can transition to
type-II, we include $\hbar ^{2}k_{z}^{2}/(2m)$ and expand the
Hamiltonian near such points, e.g., $(0,\pi /a_{y},k_{z}^{W0})$,
\begin{equation}
H(\mathbf{q})\sim (v_{0}q_{z}+v_{z}q_{z}\sigma _{z}+v_{x}q_{x}\sigma
_{x}+v_{y}q_{y}\sigma _{y})\tau _{0}
\end{equation}%
with $v_{0}=-2h_{z}/k_{L{z}}$. Remarkably, when $|h_{z}|>\hbar ^{2}k_{L{%
z}}^{2}/(4m)$ (i.e., $|v_{0}|>v_{z}$), at certain regions, the energy of
both particle bands goes negative while that of both hole bands goes
positive [as visually displayed in Fig.~\ref{Fig1}(d)], indicating that the
four-fold degenerate point becomes type-II.
Similarly, the type-I Weyl point at ${\bf k}^{W+}$ becomes type-II when $%
h_{z}<4\bar{t}-\hbar ^{2}k_{L{z}}^{2}/(4m)$ or $h_{z}>4\bar{t}+\hbar ^{2}k_{L%
{z}}^{2}/(4m)$, and the point at ${\bf k}^{W-}$ becomes type-II when $%
h_{z}>-4\bar{t}+\hbar ^{2}k_{L{z}}^{2}/(4m)$ or $h_{z}<-4\bar{t}-\hbar
^{2}k_{L{z}}^{2}/(4m)$. In Fig.~\ref{Fig2}(a), we map out the phase diagram
displaying the following phases: all degenerate points are type-I,
all are type-II, and partial type-I and partial type-II. In cold atoms,
$h_z$ can be easily tuned by changing two-photon detuning to
continuously drive the transition among different phases, thereby providing an
ideal platform to directly observe the Lifshitz transition.

We now turn to the study of surface states in this system.
In Fig.~\ref{Fig2}, the energy spectra with respect to $k_za_z$
for $k_y=0$ are plotted: (a) $h_z=0$ with type-I degenerate points
and (b) $h_z=0.4E_R$ with a type-II point.
It shows that in both cases there emerge
surface states (called Fermi arc) connecting the four-fold degenerate point
at the center with two other Weyl points on two sides. The spectra in Fig.~\ref{Fig2}(b) also
illustrate the feature of the type-II Weyl point at $%
k_za_z/\pi=-0.62$ that all the particle and hole bands near the point
at each $k_z$ are positive or negative with respect to $E_{W-}$, the energy
at the point.

Apart from the model that we have discussed, if we choose a simplified scheme
with $M_{x}=\sin (k_{L{x}}r_{x})e^{ik_{L{y}}r_{y}}$ and $M_{y}=\sin (k_{L{y}%
}r_{y})e^{-ik_{L{x}}r_{x}}$, we can still obtain both type-I and type-II
Weyl points. However, while it still respects 2D-PTRS, it breaks 2D-IS, splitting
the four-fold degenerate point into two doubly degenerate ones (see
Appendix A). To satisfy these symmetry requirements, we need to
add an additional term into the model (\ref{HTBk}) : $\tau _{z}(\alpha
_{1}\sigma _{y}-\alpha _{2}\sigma _{x})$ which respects 2D-PTRS but breaks 2D-IS.
If $V_{x}=V_{y}$ and $k_{L{x}}=k_{L{y}}$, we have $\alpha _{1}=\alpha
_{2}=\alpha $ due to a symmetry requirement (see Appendix A),
and the spectrum is $E_{\mathbf{k}}=\pm \sqrt{(d_{x}+\lambda \alpha
)^{2}+(d_{y}-\lambda \alpha )^{2}+(h_{t}-\lambda \tilde{h}_{z})^{2}}$ with $%
\lambda =\pm 1$, and Weyl points occur at $%
[k_{x}^Wa,k_{y}^Wa,k_{z}^Wa_{z}]=[\lambda \delta \theta ,\pi +\lambda \delta
\theta ,-2m\pi h_{z}/(\hbar ^{2}k_{L{z}}^{2})]$ (or $[\lambda \delta \theta
,-\lambda \delta \theta ,2m\pi (4\lambda {\bar{t}}\cos \delta \theta
-h_{z})/(\hbar ^{2}k_{L{z}}^{2})]$) with $\delta \theta =-\sin ^{-1}(\alpha
/2t_{SOx})$. Moreover, when $|h_{z}|>\hbar ^{2}k_{L{z}}^{2}/(4m) $ ($%
|h_{z}-4\lambda {\bar{t}}\cos \delta \theta |>\hbar ^{2}k_{L{z}}^{2}/(4m)$),
the former (latter) Weyl points become type-II.

\section{Realization of type-II Weyl points}

\begin{figure}[t]
\includegraphics[width=3.4in]{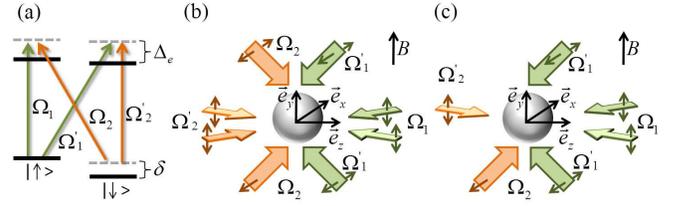}
\caption{(Color online) Schematics of laser configurations to realize the
Hamiltonian (\protect\ref{HC}) [(b)] and its simplified one [(c)]. $\Omega
_{1}$ ($\Omega _{2}$) possess the same frequency as $\Omega _{1}^\prime$ ($%
\Omega _{2}^\prime$) generated by reflecting the laser beam $%
\Omega_{1}$ ($\Omega_{2}$) by mirrors. The magnetic field is along the $y$
direction, and $\protect\delta$ is the two-photon detuning. The double
arrows denote the orientation of linear polarization of laser beams.}
\label{Fig3}
\end{figure}

To realize the Hamiltonian~(\ref{HC}), we propose an experimental scheme (as
shown in Fig.~\ref{Fig3}) that is based on a modification of the
experimental configuration in Ref.~\cite{Shuai2015}. We consider two
hyperfine states of alkali atoms such as $^{40}$K and $^{87}$Rb and employ
two independent Raman processes to create the spin-dependent optical
lattices. Each Raman process involves two linearly polarized blue-detuned
Raman laser beams with the polarization being along $y$ (parallel to the
magnetic field) and $x$, respectively. Each pair of Raman laser beams is
characterized by a pair of Rabi frequencies $[\Omega _{1}=\Omega _{10}\sin
(k_{Lx}r_{x})e^{-ik_{L{z}}r_{z}/2},\Omega _{2}=\Omega _{20}\cos (k_{L{y}%
}r_{y})e^{ik_{L{z}}r_{z}/2}]$ and $[\Omega _{1}^{\prime }=\Omega _{10}\sin
(k_{L{y}}r_{y})e^{-ik_{L{z}}r_{z}/2},\Omega _{2}^{\prime }=i\Omega _{20}\cos
(k_{L{x}}r_{x})e^{ik_{L{z}}r_{z}/2}]$, respectively. They form a standing
wave along $x$ or $y$ but remain a plane wave along $z$. Since the laser
beam $\Omega _{1}^{\prime }$ ($\Omega _{2}^{\prime }$) is obtained by
reflecting the beam $\Omega _{1}$ ($\Omega _{2}$) by mirrors, they possess
the same frequency $\omega _{1}$ ($\omega_2$), and no phase locking is
required~\cite{Xiongjun2014PRL,Yong2016PRA}. The laser beam $\Omega _{2}$
with a different frequency $\omega _{2}$ is generated by applying an
acoustic-optical modulator on a beam split from the first laser beam. Each
pair of Raman laser beams couple two hyperfine states independently, leading
to the spin-orbit coupling term $V_{SO}$ with $\Omega _{SO}=\Omega
_{10}^*\Omega _{20}/\Delta _{e}$. Moreover, owing to the stark effects,
these lasers also create optical lattices along the $x$ and $y$ directions: $%
V_{x}\sin ^{2}(k_{L{x}}r_{x})$ and $V_{y}\sin ^{2}(k_{L{y}}r_{y})$ with $%
V_{x}=V_{y}=2(|\Omega _{10}|^{2}-|\Omega _{20}|^{2})/\Delta _{e}$. The
Zeeman field $h_{z}$ is generated by the two-photon detuning $\delta $
through $h_{z}=\delta /2$ as shown in Fig. \ref{Fig3}. In comparison, a
similar laser configuration can also realize the simplified spin-orbit
coupling scheme that we mentioned before. In this case, we further simplify
the laser configuration [as shown in Fig.~\ref{Fig3}(c)] so that the Rabi
frequencies take the form $\Omega _{2}=\Omega _{20}e^{ik_{L{z}}r_{z}/2+ik_{L{%
y}}r_{y}}$ and $\Omega _{2}^{\prime }=i\Omega _{20}e^{ik{L{z}}r_{z}/2-ik_{L{x%
}}r_{x}}$, corresponding to plane waves for the second set of laser beams.

We may choose either $^{40}$K (fermions) or $^{87}$Rb (bosons) atoms for
observation of Weyl points in experiments. Here we take $^{40}$K as an
example. With a blue-detuned laser beam at wavelength $764$ nm
(corresponding to $\Delta _{e}=2\pi \times 1.38$ THz)~\cite{Jing2012PRL},
the recoil energy $E_{R}/\hbar =2\pi \times 8.5$ kHz. If we choose a
geometry with $k_{L{x}}=k_{L{y}}=k_{R}\sin \theta $ and $k_{L{z}}=2k_{R}\cos
\theta $ with $\theta =60^{\circ }$ (the angle between laser
beams and $z$ axis), $\Omega _{10}=2\pi \times 0.15$ GHz, $%
\Omega _{20}=\Omega _{10}/3$, we have $V_{x}=V_{y}=3.7E_{R}$ and $\Omega
_{SO}=0.7E_{R}$. With these experimental parameters, the tight-binding
parameters are $t_{x}=t_{y}=0.058E_{R}$, and $t_{SOx}=-t_{SOy}=0.028E_{R}$. $%
\delta $ can be readily tuned from zero and when $\delta $ crosses $0.53E_{R}
$, we will observe a Lifshitz-type quantum phase transition from type-I to
type-II Weyl points.

To detect the Weyl points of fermionic atoms and their Lifshitz phase
transition, one can measure their linear spectra along all three momenta
directions by momentum resolved radio-frequency spectroscopy, which has been
utilized for observation of a 2D Dirac cone in spin-orbit-coupled atomic
gases~\cite{Jing2015Nature,Jing2015}. The Lifshitz transition is directly
reflected by a sharp change of the dispersion of particle or hole bands
along $k_z$ near a Weyl point so that the slopes of their spectra have the
same sign. For bosonic atoms such as $^{87}$Rb, although the Fermi surface
does not exist, there is still the band structure with a touching point. One
may consider driving a BEC across a Weyl point by a constant force $F$ and
measuring the Landau-Zener tunneling probability~\cite%
{Esslinger2012Nature,Lim2012PRL,Law2015,Xiongjun2016}, which is $%
P_{LZ}=e^{-\pi E_g^2/(4vF)}$ with $E_g$ being the gap between the considered
particle and hole bands and $v$ being the velocity of the BEC~\cite{LandZene}%
. Therefore, the gap closing at a Weyl point is signalled by a peak of the
Landau-Zener tunneling probability. When the BEC bypasses a type-I (type-II)
Weyl point, a finite fraction of atoms remains in the hole band and these
atoms move in the opposite (same) direction along $z$ compared with those
tunneling into a higher band. This different behavior can be utilized to
measure the Lifshitz transition.

In summary, we have proposed a scheme well based on the current experimental
technology to realize both type-I and type-II Weyl points in the
single-particle spectra with cold atoms in an optical lattice. The proposed
system offers a unique opportunity to observe and study the topological
Lifshitz-type quantum phase transition from type-I to type-II Weyl points by
continuously tuning one of the experimental knobs.

\begin{acknowledgments}
We thank S.-T. Wang and S. A. Yang for helpful discussions. This work was supported by the
ARL, the IARPA LogiQ program, and the AFOSR MURI program.
\end{acknowledgments}

\section*{Appendix A: DERIVATION OF TIGHT-BINDING MODEL}

\setcounter{equation}{0} \setcounter{table}{0} %
\renewcommand{\theequation}{A\arabic{equation}}

In this appendix, we derive the tight-binding model from the continuous model
$H$ in Eq. (3) in the main text and compare their spectra to verity the tight-binding
model's reliability.

Let us first focus on the discretization of $H_{2D}$, which can
be written as in the second quantization language
\begin{equation}
H_{II}=\int d\mathbf{r}\hat{\psi}^{\dagger }(\mathbf{r})H_{2D}\hat{\psi}(\mathbf{r%
}),
\end{equation}%
where $\bf r$ is restricted to the $(x,y)$ plane and $%
\hat{\psi}(\mathbf{r})=[%
\begin{array}{cc}
\hat{\psi}_{\uparrow }(\mathbf{r}) & \hat{\psi}_{\downarrow }(\mathbf{r})%
\end{array}%
]^{T}$ with $\hat{\psi}_{\sigma }(\mathbf{r})$ [$\hat{\psi}_{\sigma
}^{\dagger }(\mathbf{r})$] annihilating (creating) an atom with spin $\sigma $
($\sigma =\uparrow ,\downarrow $) located at $\mathbf{r}$. They satisfy the
anti-commutation or commutation relation $[\hat{\psi}_{\sigma }(\mathbf{r}),%
\hat{\psi}_{\sigma ^{\prime }}^{\dagger }(\mathbf{r}^{\prime })]_{\pm
}=\delta _{\sigma \sigma ^{\prime }}\delta (\mathbf{r}-\mathbf{r}^{\prime })$
for fermionic atoms ($+$) or bosonic atoms ($-$), respectively. We expand the field
operator using local Wannier functions
\begin{equation}
\hat{\psi}_{\sigma }(\mathbf{r})=\sum_{n,j_x,j_y,\sigma }W_{n,j_x,j_y }\hat{c}%
_{n,j_x,j_y,\sigma },
\end{equation}%
where $W_{n,j_x,j_y}$ is the Wannier function for $\Omega_{SO}=0$ located at the site $(j_x,j_y)$ for
the $n$-th band, and $\hat{c}_{n,j_x,j_y,\sigma }$ ($\hat{c}_{n,j_x,j_y,\sigma }^\dagger$)
annihilates (creates) an atom located at the state $W_{n,j_x,j_y}$ with spin $\sigma $.
Let us focus on the physics in the lowest band and thus assume $n=1$, thereby
simplifying the above expression
\begin{equation}
\hat{\psi}_{\sigma }(\mathbf{r})\approx \sum_{j_x,j_y}W_{j_x,j_y}\hat{c}_{j_x,j_y,\sigma },
\end{equation}%
where $W_{j}=W_{j_{x}}^{x}(r_{x})W_{j_{y}}^{y}(r_{y})$
with $W_{j_{\nu }}^{\nu }(r_{\nu })=W^{\nu }(r_{\nu }-j_{\nu }a_{\nu })$
being the Wannier function along $\nu $. We note that although this is
an approximation, it proves to be qualitatively correct and we will verify it by comparing the spectra obtained by
solving the continuous model and the tight-binding one. Using this expansion, we obtain
the following tight-binding model without $V_{SO}$
\begin{eqnarray}
H_{t}=&&-\sum_{j_x,j_y,\sigma}\left(t_{x }\hat{c}_{j_x,j_y,\sigma
}^{\dagger }\hat{c}_{j_{x}+1,j_y,\sigma }
+t_{y}\hat{c}_{j_x,j_y,\sigma
}^{\dagger }\hat{c}_{j_x,j_{y}+1,\sigma } \right) \nonumber \\
&&+\text{H.c.},
\end{eqnarray}%
where $t_\nu$ with $\nu=x,y$ denote the hopping amplitudes defined as
\begin{equation}
t_{\nu } =-\int dr_{\nu }W_{j_\nu}^{\nu}\left[ \frac{p_{\nu }^{2}}{2m}+V_{\nu }\sin
^{2}(k_{L\nu }r_{\nu })\right] W_{j_{\nu }+1}^\nu.  \label{ET}
\end{equation}%
We approximately derive the tight-binding term contributed by the spin-dependent
lattice as follows
\begin{widetext}
\begin{eqnarray}
H_{SO} &=&\Omega _{SO}\int d\mathbf{r}\hat{\psi}_{\uparrow }^{\dagger }(%
\mathbf{r})\left(M_x+iM_y\right)\hat{\psi}_{\downarrow }(\mathbf{r})+\text{H.c.}, \nonumber \\
&\approx &\Omega _{SO}\sum_{j_x,j_y}\sum_{j_x^\prime,j_y^\prime}\Big [\hat{c}_{j_x,j_y,\uparrow }^{\dagger }%
\hat{c}_{j^\prime_x,j^\prime_y,\downarrow }t_{SOx}^{(j_x,j_y),(j^\prime_x,j^\prime_y)}
-i
\hat{c}_{j_x,j_y,\uparrow }^{\dagger }%
\hat{c}_{j^\prime_x,j^\prime_y,\downarrow }t_{SOy}^{(j_x,j_y),(j^\prime_x,j^\prime_y)}
\Big ]+\text{H.c.},
\end{eqnarray}%
where
\begin{eqnarray}
&&t_{SOx}^{(j_x,j_y),(j^\prime_x,j^\prime_y)}=\int d\mathbf{r}W_{j_x,j_y}\sin(k_{Lx}r_{x})\cos(k_{Ly}r_{y})W_{j^{\prime }_x,j^\prime_y}, \\
&&t_{SOy}^{(j_x,j_y),(j^\prime_x,j^\prime_y)}=-\int d\mathbf{r}W_{j_x,j_y}\cos(k_{Lx}r_{x})\sin(k_{Ly}r_{y})W_{j^{\prime }_x,j^\prime_y}.
\end{eqnarray}%
Note that we have added a minus sign in the definition of $t_{SOy}$ in order to write the
Hamiltonian in a compact form.

\begin{figure*}[t]
\includegraphics[width=5.2in]{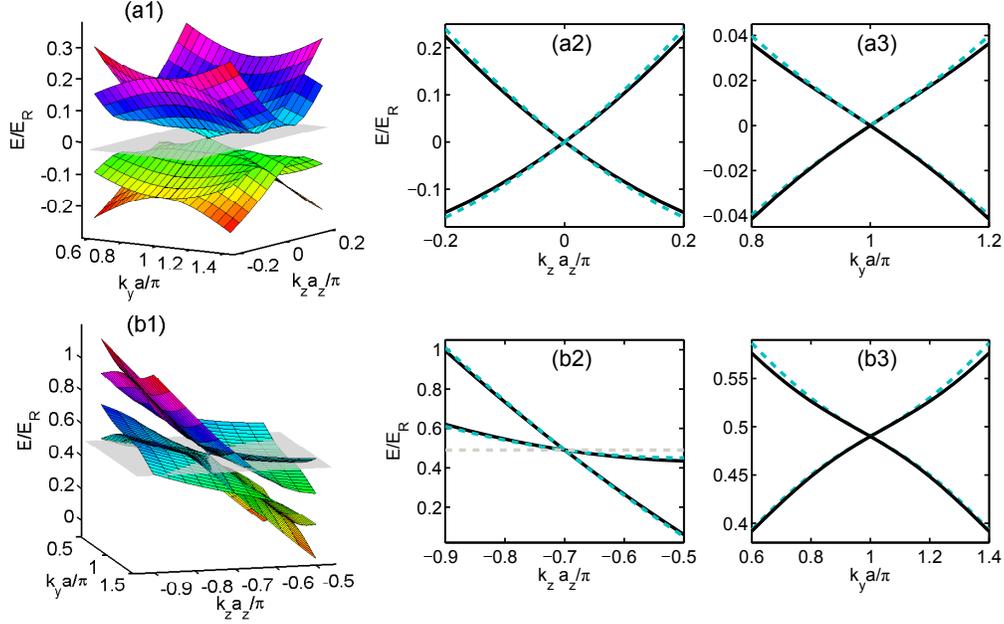}
\caption{(Color online) Single-particle spectra in the ($k_y,k_z$) plane for $k_x=0$ obtained
by ab initio theory (i.e., diagonalizing the continuous model) in (a1) and (b1) [the spectra
of tight-binding model are plotted in Fig. 1(c) and (d) in the main text].
To compare these spectra in detail, we further plot them (with solid black and dashed cyan
lines denoting the spectra of continuous and tight-binding model, respectively) around a degenerate
point [$(k_y a/\pi=1,k_z=0)$ for the first row panel and $(k_y a/\pi=1,k_z a_z/\pi=-0.7)$
for the second row panel] as a function of $k_z$ and $k_y$ in (a2,b2) and (a3,b3), respectively.
Here, we choose $V_x=V_y=3.7E_{R}$ and $\Omega_{SO}=0.7E_{R}$ corresponding to
the tight-binding model with $t_x=t_y=0.058E_{R_x}$ and $t_{SO}=0.028E_{R}$.
For (a1-a3), $h_z=0$, while for (b1-b3), $h_z=0.7E_R$. For the
spectra of the continuous model, we have shifted the energy at the degenerate
point to the energy at the same point of the spectra of the tight-binding model.}
\label{SIfig1}
\end{figure*}

\begin{figure*}[t]
\includegraphics[width=7in]{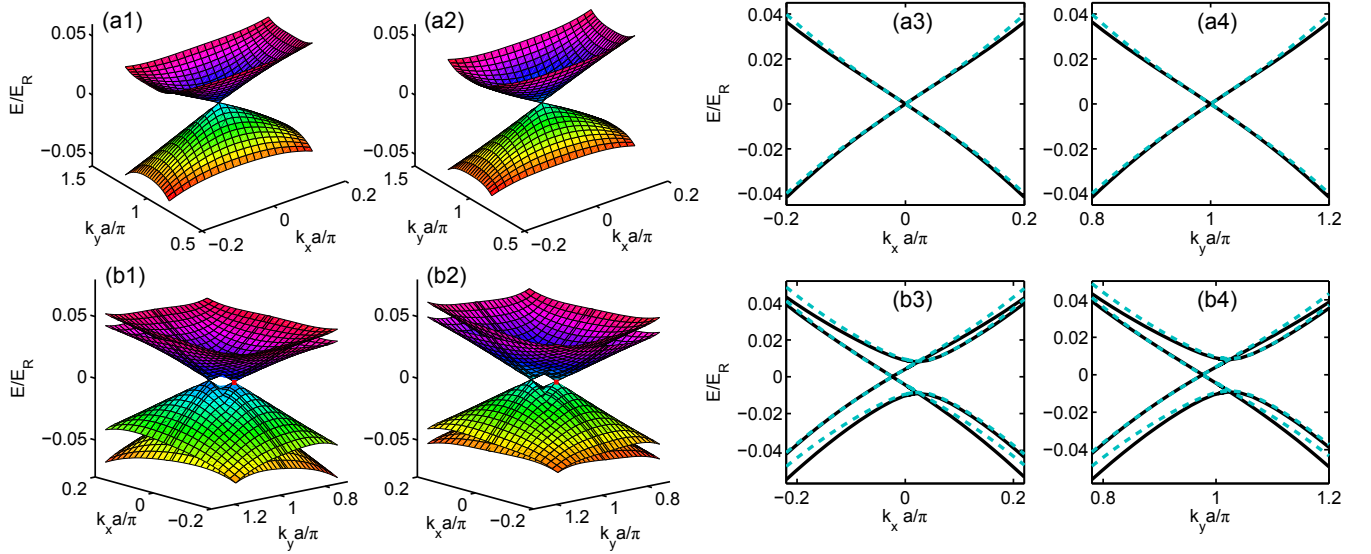}
\caption{(Color online)
Single-particle spectra in the ($k_x,k_y$) plane for $k_z=0$, obtained
by ab initio theory in (a1) and (b1) and
by diagonalizing the tight-binding model in (a2) and (b2).
(a1-a4) correspond to the model
with $M_x=\sin(k_{Lx}r_{x})\cos(k_{Ly}r_{y})$,
$M_y=\sin(k_{Ly}r_{y})\cos (k_{Lx}r_{x})$, and (b1-b4)
to the simplified model with $M_x=\sin(k_{Lx}r_x)e^{ik_{Ly}r_y}$
and $M_y=\sin(k_{Ly}r_y)e^{-ik_{Lx}r_x}$.
To compare these spectra in detail, we further plot them (with solid black and dashed cyan
lines denoting the spectra of continuous and tight-binding model, respectively) around a degenerate
point [$(k_x=0,k_y a/\pi=1)$ for the first row panel and $(k_x a/\pi=-0.024,k_y a/\pi=0.976)$ denoted by the red square
in (b1) and (b2) for the second row panel] as a function of $k_x$ and $k_y$ in (a3,b3) and (a4,b4), respectively.
Here, we choose $V_x=V_y=3.7E_{R}$, $\Omega_{SO}=0.7E_{R}$, and $h_z=0$, corresponding to the tight-binding model
with $t_x=t_y=0.058E_{R_x}$ and $t_{SO}=0.028E_{R}$. For comparison, we have shifted the
spectra of the continuous model so that the energy at the degenerate point is zero.}
\label{SIfig2}
\end{figure*}

Employing the condition $W_{0}^{\nu }(r_{\nu })=W_{0}^{\nu }(-r_{\nu })$
given that one of the optical wells is located at $\mathbf{r}=(0,0)$ when $\Omega_{SO}=0$, we get
\begin{eqnarray}
&&t_{SOx}^{(j_x,j_y),(j_x,j_y^\prime)}=t_{SOy}^{(j_x,j_y),(j_x^\prime,j_y)}=0, \\
&&t_{SOx}^{(j_x,j_y),(j_x+1,j_y)}=-t_{SOx}^{(j_x,j_y),(j_x-1,j_y)}=(-1)^{j_x+j_y}t_{SOx}^{(0,0),(1,0)}, \\
&&t_{SOy}^{(j_x,j_y),(j_x,j_y+1)}=-t_{SOy}^{(j_x,j_y),(j_x,j_y-1)}=(-1)^{j_x+j_y}t_{SOy}^{(0,0),(0,1)},
\end{eqnarray}%
where the last two relations are obtained because $\sin(k_{L\nu}(r_{\nu}+a_\nu))=-\sin(k_{L\nu}r_{\nu})$ and $\cos(k_{L\nu}(r_{\nu}+a_\nu))=-\cos(k_{L\nu}r_{\nu})$. Therefore, if we only consider the nearest-neighbor
hopping, we obtain the following spin-orbit coupling term of the tight-binding model:
\begin{eqnarray}
H_{SO} &\approx &\Omega _{SO}\sum_{j_x,j_y}\left(\hat{c}_{j_x,j_y,\uparrow }^{\dagger }%
\hat{c}_{j_x+1,j_y,\downarrow }t_{SOx}^{(j_x,j_y),(j_x+1,j_y)}
+\hat{c}_{j_x,j_y,\uparrow }^{\dagger }%
\hat{c}_{j_x-1,j_y,\downarrow }t_{SOx}^{(j_x,j_y),(j_x-1,j_y)} \right. \nonumber \\
&&\left.-i\hat{c}_{j_x,j_y,\uparrow }^{\dagger }%
\hat{c}_{j_x,j_y+1,\downarrow }t_{SOy}^{(j_x,j_y),(j_x,j_y+1)}
-i\hat{c}_{j_x,j_y,\uparrow }^{\dagger }%
\hat{c}_{j_x,j_y-1,\downarrow }t_{SOy}^{(j_x,j_y),(j_x,j_y-1)}
\right)+\text{H.c.},\\
&=&\Omega _{SO}\sum_{j_x,j_y}(-1)^{j_x+j_y}\left[t_{SOx}^{(0,0),(1,0)}(\hat{c}_{j_x,j_y,\uparrow }^{\dagger }%
\hat{c}_{j_x+1,j_y,\downarrow }
-\hat{c}_{j_x,j_y,\uparrow }^{\dagger }%
\hat{c}_{j_x-1,j_y,\downarrow })\right. \nonumber \\
&&\left.-it_{SOy}^{(0,0),(0,1)}(\hat{c}_{j_x,j_y,\uparrow }^{\dagger }%
\hat{c}_{j_x,j_y+1,\downarrow }
-\hat{c}_{j_x,j_y,\uparrow }^{\dagger }%
\hat{c}_{j_x,j_y-1,\downarrow }
)\right]+\text{H.c.},\\
&=&\sum_{\bf x}\sum_{\nu=x,y}(-1)^{j_x+j_y}t_{SO\nu}\hat{c}_{{\bf x}}^{\dagger }\sigma_\nu\hat{c}_{{\bf x}+{\bf g}_\nu}+\text{H.c.},
\end{eqnarray}
where in the last step we have recast the Hamiltonian into a compact form by defining
$t_{SO\nu}=\Omega_{SO}t_{SO\nu}^{(0,0),(0,1)}$, ${\bf g}_\nu=a_\nu{\bf e}_\nu$, $\hat{c}^\dagger_{{\bf x}}=(
\begin{array}{cc} \hat{c}^\dagger_{{\bf x},\uparrow} & \hat{c}^\dagger_{{\bf x},\downarrow}
\end{array})$ with $\hat{c}^\dagger_{{\bf x},\sigma}\equiv\hat{c}^\dagger_{j_x,j_y,\sigma}$ and
${\bf x}=j_{x}a_x{\bf e}_x+j_{y}a_y{\bf e}_y$. When $V_x=V_y$ and $k_{Lx}=k_{Ly}$, $t_{SOx}=-t_{SOy}$.
In 3D, after replacing $\hat{c}_{{\bf x},\sigma}$ with $\hat{c}_{k_z,{\bf x},\sigma}$ and
including $H_t$ and the Zeeman field term, we obtain $H_{TB}$ in Eq.(5) in the main text.

\end{widetext}

To verify the reliability of the tight-binding model, in Fig.~\ref{SIfig1} and Fig.~\ref{SIfig2},
we compare the energy spectra of the tight-binding model with those
of the continuous model, which are numerically calculated using Fourier
series expansion of a Bloch function. In Fig.~\ref{SIfig1} (a1) and (b1),
we present the spectra of the continuous model, which qualitatively agree with
their tight-binding counterparts in Fig. 1(c) and (d) of the main text. To see their
difference more quantitatively, we plot
their spectra around a degenerate point with respect to $k_z$ and $k_y$
in Fig.~\ref{SIfig1}(a2,b2) and (a3,b3), respectively, illustrating excellent agreement between
these two approaches in the vicinity of a degenerate point; this implies that the tight-binding
model can well characterize these degenerate points. As momenta move far away from
these points, there appears a slight discrepancy, which might be reduced when
the next-nearest-neighbor hopping is included. In Fig.~\ref{SIfig2}(a1-a2),
we further compare their spectra in the $(k_x,k_y)$ plane; the doubly degenerate spectra
in Fig.~\ref{SIfig2}(a) and those in Fig.~\ref{SIfig2}(b) are in qualitative
agreement with each other and both exhibit a four-fold degenerate point. In addition,
a more quantitative comparison in Fig.~\ref{SIfig2}(a3) and (a4) demonstrates the reliability
of the tight-binding model.

For the
simplified model with $M_x=\sin(k_{Lx}r_x)e^{ik_{Ly}r_y}$
and $M_y=\sin(k_{Ly}r_y)e^{-ik_{Lx}r_x}$,
it breaks 2D-IS but still respects 2D-PTRS and therefore in the tight-binding model
we need to add an additional term $\tau_z(\alpha_1\sigma_y-\alpha_2\sigma_x)$ that
preserves 2D-PTRS but lacks 2D-IS. When $V_x=V_y$ and $k_{Lx}=k_{Ly}$,
we have $\alpha_1=\alpha_2=\alpha$ because the 2D continuous Hamiltonian
respects a $\Pi_C$ symmetry, i.e., $\Pi_CH_{2D}\Pi_C^{-1}=H_{2D}$, where $\Pi_C=M\sigma_xS_4^{-1}$ with $MH_{2D}(x,y)M^{-1}=H_{2D}(y,x)$;
the representation of this symmetry in the momentum space corresponds to $\Pi H_{TB}^{2D}({\bf k})\Pi^{-1}=H_{TB}^{2D}(k_y,k_x)$
[$H_{TB}^{2D}\equiv H_{TB}(\tilde{h}_z=0)$] with $\Pi=(\sigma_x-\sigma_y)/\sqrt{2}$, indicating that the additional term must take the form of $\sigma_x-\sigma_y$. In Fig.~\ref{SIfig2}(b1-b2), we plot the spectra of the continuous and tight-binding models and both figures illustrate
that a four-fold degenerate touching point splits into two doubly degenerate ones. Our further comparison around a touching point
in Fig.~\ref{SIfig2}(b3-b4) shows the quantitative agreement of the latter with the former.

\section*{Appendix B: ANALYSIS OF MANY-BODY EFFECTS}
In this section, we make a crude estimate of many-body effects on the degenerate points in the presence of
weak repulsive atom-atom interactions for fermionic atoms. For alkali atoms, the interaction readily tuned
by Feshbach resonances is short-range and can be written
as (we only consider dominant on-site interactions)
\begin{equation}
H_{Int} =g\sum_{k_{z}k_{z}^{\prime}Q}\sum_{{\bf x}}\hat{c}_{k_{z}{\bf x}\uparrow}^{\dagger}\hat{c}_{-k_{z}+Q{\bf x}\downarrow}^{\dagger}\hat{c}_{-k_{z}^{\prime}+Q{\bf x}\downarrow}\hat{c}_{k_{z}^{\prime}{\bf x}\uparrow},
\end{equation}
where $g$ denotes the strength of interactions proportional to the $s$-wave scattering length. For weak
interactions, using mean-field approximations yields
\begin{eqnarray}
H_{Int}\approx &&\sum_{{\bf k}}{\Psi}({\bf k})^{\dagger}H_{Int}^M({\bf k}){\Psi}({\bf k})+N\sum_{D=A,B}\frac{|m_{\parallel,D}|^{2}}{g} \nonumber \\
&&-N\sum_{D=A,B}\frac{m_{z,\uparrow,D}m_{z,\downarrow,D}}{g},
\end{eqnarray}
where $N$ is the number of sites in the $(x,y)$ plane,
\begin{eqnarray}
m_{z,\sigma,D}  &=&g\sum_{k_{z}}\langle\hat{D}_{k_{z}{\bf x}\sigma}^{\dagger}\hat{D}_{k_{z}{\bf x}\sigma}\rangle, \\
m_{\parallel,D} &=&m_{x,D}+im_{y,D} \nonumber \\
&=&g\sum_{k_{z}}\langle\hat{D}_{k_{z}{\bf x},\uparrow}^{\dagger}\hat{D}_{k_{z}{\bf x},\downarrow}\rangle
\end{eqnarray}
and
\begin{eqnarray}
H_{Int}^{M}({\bf k})=&&m_{z,1}+m_{z,2}\sigma_{z}+m_{x,1}\sigma_{x}+m_{y,1}\sigma_{y}+m_{z,3}\tau_{z} \nonumber \\
&&+m_{z,4}\sigma_{z}\tau_{z}+m_{x,2}\sigma_{x}\tau_{z}+m_{y,2}\sigma_{y}\tau_{z},
\end{eqnarray}
with
\begin{eqnarray}
m_{z,1} =\frac{1}{4}(m_{z,\downarrow,A}+m_{z,\uparrow,A}+m_{z,\downarrow,B}+m_{z,\uparrow,B}), \\
m_{z,2} =\frac{1}{4}(m_{z,\downarrow,A}-m_{z,\uparrow,A}+m_{z,\downarrow,B}-m_{z,\uparrow,B}), \\
m_{z,3} =\frac{1}{4}(m_{z,\downarrow,A}+m_{z,\uparrow,A}-m_{z,\downarrow,B}-m_{z,\uparrow,B}), \\
m_{z,4} =\frac{1}{4}(m_{z,\downarrow,A}-m_{z,\uparrow,A}-m_{z,\downarrow,B}+m_{z,\uparrow,B}),
\end{eqnarray}
and
\begin{eqnarray}
m_{x,1} =-\frac{1}{2}(m_{x,A}+m_{x,B}), \\
m_{x,2} =-\frac{1}{2}(m_{x,A}-m_{x,B}), \\
m_{y,1} =-\frac{1}{2}(m_{y,A}+m_{y,B}), \\
m_{y,2} =-\frac{1}{2}(m_{y,A}-m_{y,B}).
\end{eqnarray}
If we consider using the ground state of non-interacting fermionic
atoms as the initial state for iteration while searching for the many-body
ground state, we have $m_{z,\sigma,A}=m_{z,\sigma,B}$ and $m_{x,A}=m_{x,B}=m_{y,A}=m_{y,B}=0$,
so that $m_{z,3}=m_{z,4}=m_{x,1}=m_{y,1}=m_{x,2}=m_{y,2}=0$.
Based on this argument, we have
\begin{equation}
H_{Int}({\bf k})=m_{z,1}+m_{z,2}\sigma_{z}.
\end{equation}
Clearly, the presence of interactions may induce an effective Zeeman field,
which will shift the locations of degenerate points along the z direction and
may cause the transition between the type-I and type-II, but will neither destroy
Weyl nor four-fold degenerate points.

\end{document}